\documentclass[]{spiejour}  
\usepackage[]{graphicx}
\usepackage[]{amsmath}
\usepackage[]{bm}

\title{Magnetically operated nanorelay based on two single-walled carbon nanotubes filled with endofullerenes Fe@C$_\text{20}$} 

\author{Nikolai~A. Poklonski,\supscr{a} Eugene~F. Kislyakov,\supscr{a} Sergey~A. Vyrko,\supscr{a} Nguyen~Ngoc~Hieu,\supscr{b} Oleg~N.~Bubel',\supscr{a} Andrei~I. Siahlo,\supscr{a} Irina~V.~Lebedeva,\supscr{c} Andrey~A. Knizhnik,\supscr{d,e} Andrey~M.~Popov,\supscr{f} and Yurii~E.~Lozovik\supscr{f}}

\affiliation{\supscr{a}Belarusian State University, pr. Nezavisimosti 4, Minsk 220030, Belarus\\
\linkable{poklonski@bsu.by} \\
\supscr{b}Institute of Physics and Electronics, Hanoi, Vietnam\\
\supscr{c}Moscow Institute of Physics and Technology, Institutskii per., 9, Dolgoprudny 141701, Moscow Region, Russia\\
\supscr{d}Kintech Lab Ltd, Kurchatov Sq. 1, Moscow 123182, Russia\\
\supscr{e}RRC ``Kurchatov Institute'', Kurchatov Sq. 1, Moscow 123182, Russia \\
\supscr{f}Institute of Spectroscopy, Troitsk 142190, Moscow Region, Russia\\
\linkable{am-popov@isan.troitsk.ru}, \linkable{lozovik@isan.troitsk.ru\llap{\phantom{p}}}
}

\renewcommand{\vec}[1]{\bm{\mathrm{#1}}}     
 
\begin{document} 
  \maketitle 

\hyphenation{nano-tu-be nano-tu-bes semi-empi-ri-cal endo-fulle-re-nes endo-fulle-re-ne nano-re-lay}

\begin{abstract}
Structural and energy characteristics of the smallest magnetic endofullerene Fe@C$_\text{20}$ were calculated using the density functional theory. The ground state of Fe@C$_\text{20}$ was found to be a septet state, and the magnetic moment of Fe@C$_\text{20}$ was estimated to be 8 Bohr magnetons. The characteristics of an (8,8) carbon nanotube with a single Fe@C$_\text{20}$ inside were studied with a semiempirical approach. The scheme of a magnetic nanorelay based on cantilevered nanotubes filled with magnetic endofullerenes was examined. This nanorelay is turned on as a result of bending of nanotubes by a magnetic force. The operational characteristics of such a nanorelay based on (8,8) and (21,21) nanotubes fully filled with Fe@C$_\text{20}$ were estimated and compared to the ones of a nanorelay made of a (21,21) nanotube fully filled with experimentally observed (Ho$_\text{3}$N)@C$_\text{80}$ with the magnetic moment of 21 Bohr magnetons. The room-temperature operation of (21,21) nanotube-based nanorelays was demonstrated.
\end{abstract}


\section{Introduction}
\label{sect:intro}  
The remarkable elastic properties and metallic conductivity of carbon nanotubes (CNTs) allow applications of nanotubes as parts of nanoelectromechanical systems (see, e.g., Refs.~\citen{Dong07, Bichoutskaia07, Lozovik07}). In fact, a wide set of electromechanical nanorelays based on the crossbar relative position of two carbon nanotubes~\cite{Rueckes00}, cantilevered~\cite{Lee04, Axelsson05, Jang05, Jang08} and suspended~\cite{Dujardin05} nanotubes, telescopic extension of nanotubes~\cite{Deshpande06} and a magnetic shuttle inside a carbon nanotube~\cite{Begtrup09} has been implemented. Operational principles of these nanorelays and possibility to use them as memory cells are considered in Ref.~\citen{Bichoutskaia08}. 

A variety of endofullerenes and nanotubes filled with fullerenes (nanotube peapods) are obtained in macroscopic amounts \cite{Kitaura07}, including nanotube peapods filled with magnetic endofullerenes~\cite{Hirahara00, Shiozawa06, Sun04, Kitaura07PRB172409}. In the present work we propose a new type of a nanorelay based on nanotube peapods encapsulating magnetic endofullerenes. Operation of this new nanorelay is determined by the balance of three major forces: magnetic, elastostatic, and van der Waals ones.

Endofullerenes with encapsulated iron atoms were observed experimentally~\cite{Pradeep92, Churilov97}.
Recently the magnetic moment of the endofullerene Fe$_\text{3}$@C$_\text{60}$ was calculated~\cite{Gao08}. Here we consider the fullerene C$_\text{20}$ with encapsulated iron atom as the simplest model of a magnetic endofullerene. We show the possibility that the smallest magnetic endofullerene Fe@C$_\text{20}$ can exist. The density functional theory (DFT) approach is used here to study the structure, energetics and the magnetic moment of the endofullerene Fe@C$_\text{20}$. The possibility of encapsulation of small fullerenes inside nanotubes was studied in Ref.~\citen{Zhou06}. It was shown that an armchair (8,8) nanotube is the smallest carbon nanotube (CNT) which can encapsulate fullerene C$_\text{20}$~\cite{Zhou06}. In the present work the binding energy, the equilibrium position, and energetic barriers for motion and rotation of the endofullerene Fe@C$_\text{20}$ inside the (8,8) nanotube are calculated using a semiempirical potential.

The magnetic moment was measured for a wide set of fullerenes with encapsulated magnetic atoms~\cite{Kitaura07PRB172409, Huang00, Bondino06, Wolf05, Okimoto08, Smirnova08, Tiwari08}. It was also shown that the magnetic moment is larger for magnetic endofullerenes inside carbon nanotubes than for the same isolated magnetic endofullerenes~\cite{Kitaura07PRB172409}. Nowadays the largest value of the magnetic moment of 21 Bohr magnetons was observed for (Ho$_\text{3}$N)@C$_\text{80}$~\cite{Wolf05}. Here operational characteristics of the proposed magnetic nanorelay are calculated for the cases of nanotubes filled with the smallest theoretically possible magnetic endofullerene Fe@C$_\text{20}$ and the magnetic endofullerene (Ho$_\text{3}$N)@C$_\text{80}$ with the largest observed magnetic moment.

The paper is organized as follows. Section 2 presents the DFT calculations of properties of the endofullerene Fe@C$_\text{20}$. Section 3 is devoted to characteristics of endofullerene encapsulation inside the (8,8) carbon nanotube. Section 4 contains the principal scheme of the nanotube peapod-based nanorelay controlled by magnetic field and estimations of its operational characteristics. Our conclusions are summarized in Sec.~5.

\section{Endofullerene F\MakeLowercase{e}@C$_\text{20}$}
\label{sect:02}  

The study of carbon-iron systems is important for physics and chemistry of carbon nanostructures.
It is known that iron is a catalyst for synthesis of carbon nanostructures and diamond. Thus, in the process of CNT synthesis~\cite{Charlier01, Joselevich08} iron nanoparticles were sometimes left on free tips of nanotubes. The mechanisms of such processes are still not established.
We consider behavior of an iron atom in a quantum-size structure, fullerene C$_\text{20}$, as an example of one of the simplest carbon-iron systems.

According to the quantum chemical calculations~\cite{Poklonskij02ZhPS283}, C atoms of the fullerene C$_\text{20}$ are located at 0.205~nm from its center (for the average C--C bond length of 0.146~nm). The iron atom in the ferrocene C$_\text{10}$H$_\text{10}$Fe is located at the same distance from the carbon atoms (see, e.g.,~\cite{Luthi80}). Due to this fact we assume that an iron atom can be put inside the fullerene C$_\text{20}$. The iron atom in the center of the fullerene C$_\text{20}$ is located in the force field with the approximate symmetry of icosahedron $\bm{I}_h$. The field with such symmetry does not split the $D$ terms of atoms~\cite{Inui96, Poklonskij03gruppy}, therefore the iron atom can preserve its angular momentum inside the fullerene~C$_\text{20}$.

Spin states of the endofullerene Fe@C$_\text{20}$ have been calculated using the spin-polarized density functional theory implemented in the NWChem 4.5 code~\cite{NWChem03} with the Becke--Lee--Yang--Parr exchange-correlation functional (B3LYP)~\cite{Becke93, Lee88}. 18 inner electrons of the iron atom are emulated with the help of the effective core potential~--- CRENBS ECP~\cite{Ross90} (only 8 valence \emph{s}--\emph{d} electrons were taken into account explicitly). The 6-31G* basis set is used to describe electrons of the carbon atoms.

The calculated energies of the quintet ($S = 2$), triplet ($S = 1$) and singlet ($S = 0$) states of the endofullerene Fe@C$_\text{20}$ are found to be 0.61, 0.86 and 1.21~eV higher than the energy of the ground state.
The ground state of the endofullerene Fe@C$_\text{20}$ is found to be a septet state (with the total spin moment $S = 3$ and the multiplicity $2S + 1 = 7$) and has the $\bm{C}_{2h}$ symmetry. Since the ground state $^5D_4$ of a free iron atom~\cite{Emsli93} has the total moment $J = 4$ and the multiplicity $2S + 1 = 5$, the atomic magnetic moment of the iron atom inside the fullerene C$_\text{20}$ brings about the ferromagnetic ordering of $\pi$-\hspace{0pt}electrons of the unfilled electron shell of the fullerene.
The magnetic moment of an iron atom is $\vec{M}_\text{Fe} = -g\vec{J}\mu_\text{B}$, where $\mu_\text{B} = \text{5.788}{\cdot}10^{-5}$~meV$/$mT is the Bohr magneton and $g = 1 + [J(J + 1) - L(L + 1) + S(S + 1)]/[2J(J + 1)]$ is the Lande splitting factor (spectroscopic splitting factor), $\vec{J}$ is the total angular momentum of an iron atom. For the ground state $^5D_4$ term of the iron atom we use~\cite{Emsli93}: $J = 4$, $L = 2$, $S = 2$ and the Lande factor $g = 3/2$, this yields $M_\text{Fe} = g\sqrt{J(J + 1)}\mu_\text{B} \approx 6\mu_\text{B}$. We assume that the endofullerene Fe@C$_\text{20}$ magnetic moment is made up of the iron atom magnetic moment and the spin magnetic moment of the carbon cage ($M_\text{cc} \approx 2\mu_\text{B}$
). So we obtain $M_\text{ef} \approx M_\text{Fe} + M_\text{cc} = \text{8}\mu_\text{B}$.

\begin{figure}
   \begin{center}
   \begin{tabular}{c}
   \includegraphics{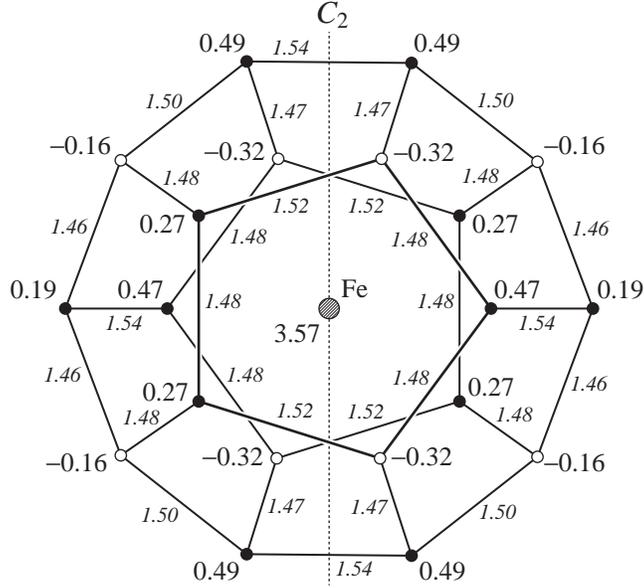}
   \end{tabular}
   \end{center}
   \caption 
   { \label{fig:01} 
The calculated structure and spin density of atoms in units of $\hbar/2$ of the ground state of the endofullerene Fe@C$_\text{20}$. The lengths of bonds (in \aa ngstr\"oms) are in italics. The dotted line $C_2$ represents the two-fold symmetry axis.} 
   \end{figure}

The calculated structure of the ground state of the endofullerene Fe@C$_\text{20}$ is shown in Fig.~\ref{fig:01}. The iron atom is located in the center of the endofullerene. The distances between the iron atom and the carbon atoms are 0.206 and 0.210~nm for the carbon atoms shown in Fig.~\ref{fig:01} by the open and filled circles, respectively. Note that these distances are smaller for the carbon atoms with anti-parallel to the iron magnetic moment spin density (attracting to the iron atom) and larger for the carbon atoms with parallel spin density (repulsing from the iron atom).

The Mulliken population analysis~\cite{Mulliken55} is used to calculate the charge distribution of Fe@C$_\text{20}$. It is found that valence electrons of the iron atom partially shift to the carbon cage, and the iron atom gains the charge of 1.2$e$, while the carbon atoms located closer to and farther from the iron atom gain the charges of $-$0.04$e$ and $-$0.07$e$, respectively, where $e$ is the elementary charge.

Our calculations of the empty fullerene C$_\text{20}$ show that it is a fairly stable system with the binding energy of 6.15~eV$/$atom (slightly less than 7.0~eV$/$atom~\cite{Tomanek91} for the fullerene C$_\text{60}$) and the singlet ground state (the triplet state is 65~meV higher). Though, isomers of cluster C$_\text{20}$ with lower energy may exist, calculations~\cite{Davydov05} show the high thermal stability of the fullerene C$_\text{20}$. Up to now the fullerene C$_\text{20}$ was detected only in the gas phase~\cite{Prinzbach00}. According to our calculations the incorporation of an iron atom into the fullerene C$_\text{20}$ requires 5.1~eV. This is less than the C atom binding energy in the empty fullerene C$_\text{20}$, which indicates the possibility of Fe@C$_\text{20}$ existence, while the stability of such a system remains at the same level as for the empty fullerene C$_\text{20}$. 

\section{F\MakeLowercase{e}@C$_\text{20}$ inside CNT (8,8)}
\label{sect:03}  

The structure of the endofullerene Fe@C$_\text{20}$ described in Sec.~2 is used to find the ground state and to study motion of this endofullerene inside the (8,8) carbon nanotube with all equal bonds 0.1423~nm in length. The van der Waals interaction between carbon atoms of the endofullerene and the nanotube at the interatomic distance $r$ is described by the Lennard-Jones 12--6 potential
\begin{equation}\label{eq:01}
   U_\text{W} = 4\varepsilon \biggl[\biggl(\frac{\sigma}{r}\biggr)^{\!12} - \biggl(\frac{\sigma}{r}\biggr)^{\!6}\,\biggr]
\end{equation}
with the parameters $\varepsilon = \text{2.755}$~meV and $\sigma = \text{0.3452}$~nm. These parameters of the Lennard-\hspace{0pt}Jones potential for the fullerene-nanotube interaction are obtained as the average values of the parameters~\cite{Girifalco00} for fullerene-fullerene and fullerene-graphene interactions, in accordance with the procedure described in~\cite{Girifalco00}. The cut-off distance $r = r_\text{c}$ of the Lennard-Jones potential is taken equal to $r_\text{c} ={}$1.5~nm. For this cut-off distance the errors of our calculation in the interaction energy $U_\text{W}$ between the endofullerene Fe@C$_\text{20}$ and the (8,8) nanotube and in the barriers for relative motion and rotation of the endofullerene inside the nanotube are less than 0.1\%.
Both the fullerene and the nanotube are considered to be rigid. Influence of structure deformation is not essential both for the interwall interaction of carbon nanotubes~\cite{Kolmogorov00, Belikov04} and the intershell interaction of carbon nanoparticles~\cite{Lozovik00, Lozovik02}. For example, taking into consideration the structure deformation of the shells of the C$_\text{60}$@C$_\text{240}$ nanoparticle only slight changes of the barriers for relative rotation of the shells are found (less than 1\%)~\cite{Lozovik00, Lozovik02}.
   \begin{figure}[!b]
   \begin{center}
   \begin{tabular}{c}
   \includegraphics{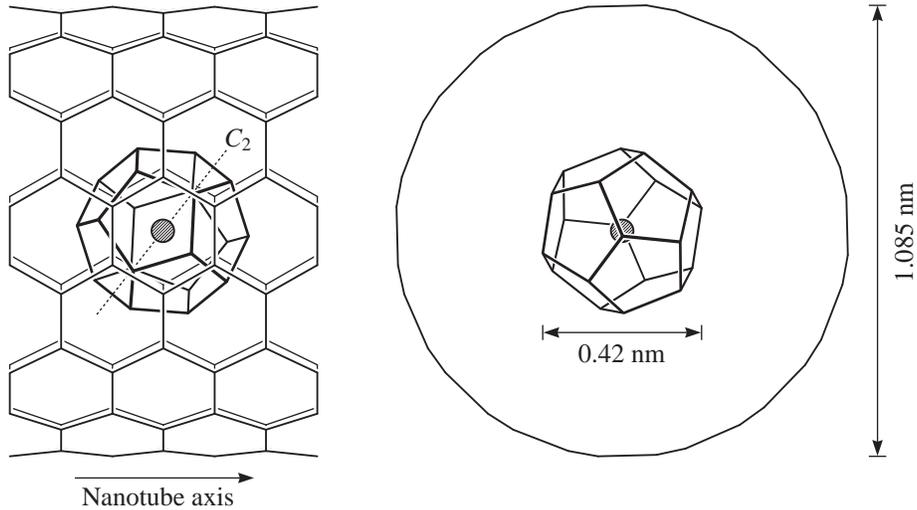}
   \end{tabular}
   \end{center}
   \caption 
   { \label{fig:02} 
The calculated structure of the ground state of the (Fe@C$_\text{20}$)@CNT(8,8) peapod nanotube. The diameters of the endofullerene and the nanotube are indicated. Thinner lines correspond to the back side.}
   \end{figure} 

   \begin{figure}
   \begin{center}
   \begin{tabular}{c}
   \includegraphics{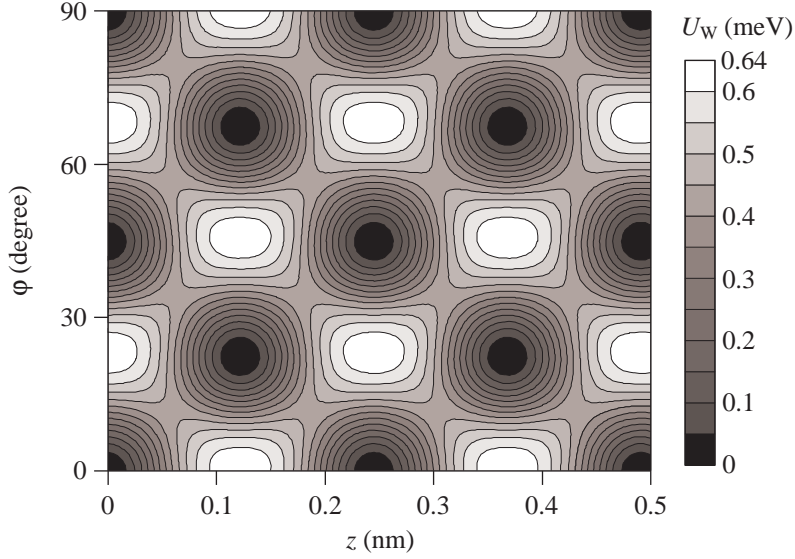}
   \end{tabular}
   \end{center}
   \caption 
   { \label{fig:03} 
The contour map of the interaction energy $U_\text{W}(\varphi, z)$ between the endofullerene Fe@C$_\text{20}$ and the (8,8) carbon nanotube as a function of the relative displacement $z$ (in nm) of Fe@C$_\text{20}$ along the nanotube axis and the angle $\varphi$ (in degrees) of relative rotation of Fe@C$_\text{20}$ about it. The equipotential lines are drawn at an interval of 0.05~meV. The maximum $U_\text{W}$ value is 0.64~meV.} 
   \end{figure} 

The ground state interaction energy between the endofullerene Fe@C$_\text{20}$ and the (8,8) nanotube is calculated to be $-$1.598~eV. The position of the endofullerene Fe@C$_\text{20}$ inside the (8,8) nanotube at the ground state is shown in Fig.~\ref{fig:02}, displaying the angle of 52$^{\circ}$ between the symmetry axes of the endofullerene and the nanotube. The contour map of the interaction energy $U_\text{W}(\varphi, z)$ is presented in Fig.~\ref{fig:03}, where $\varphi$ is the angle of relative rotation of Fe@C$_\text{20}$ about the nanotube axis, and $z$ is the relative displacement of Fe@C$_\text{20}$ along this axis. Figure~\ref{fig:03} shows that both barriers for motion of the endofullerene along the nanotube axis and for rotation of endofullerene about this axis are 0.64~meV. 
The minimal barrier for motion of endofullerene along the nanotube axis corresponds to the case where this motion is accompanied by rotation about the nanotube axis. Figure~\ref{fig:03} shows that the value of this minimal barrier between adjacent minima is 0.45~meV.

The calculations performed here show that free motion of the endofullerene along the nanotube axis occurs at room temperature. Therefore a single endofullerene Fe@C$_\text{20}$ is able to move to a nanotube end where it can be trapped in the potential well associated with the interaction between the endofullerene and the nanotube cap. In much the same way  endofullerenes are able to assemble in a one-\hspace{0pt}dimensional cluster inside the nanotube due to van der Waals interaction between the endofullerenes.

\section{Nanotube-based magnetically operated nanorelay}
\label{sect:04}  

\begin{figure}
   \begin{center}
   \begin{tabular}{ccc}
   ~~\,\includegraphics{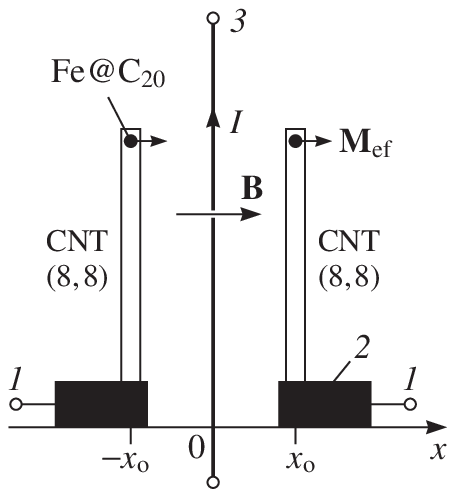} &~~ &
   ~~~~~\includegraphics{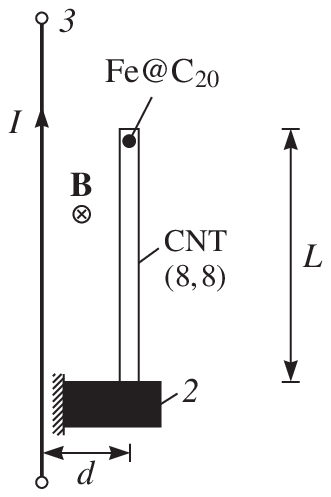} \\
   (a) & & (b)
   \end{tabular}
   \end{center}
   \caption 
   { \label{fig:04} 
The scheme of the magnetic nanorelay based on nanotubes: \emph{1} is the controlled circuit, \emph{2}~is the electrode with the attached peapod nanotube, \emph{3} is the wire passing the current $I$ through it and used for inducing the magnetic field with the induction $B$, which brings tips of nanotubes into contact with each other.} 
   \end{figure}

We propose a new type of a nanorelay based on peapod nanotubes with encapsulated magnetic endofullerenes (see Fig.~\ref{fig:04}) which allows to disconnect control and controlled circuits, analogously to known~\cite{Peek55} macroscopic reed switches. The proposed nanorelay has the following operational principles. 
When a magnetic field is applied, the majority of the magnetic moments of the endofullerenes lines up in the direction of the magnetic field as it is shown in Fig.~\ref{fig:04}(a). As a result of the magnetic moment alignment, the attraction between the nanotubes arises. The condition for the nanorelay switching is determined by a balance of the magnetic and elastostatic forces.

The interaction energy of two magnetic dipoles with the moments $\vec{M}_i$ and $\vec{M}_j$ of the endofullerenes encapsulated in the right and left nanotubes, respectively, is given by (see, e.g., Ref.~\citen{White07})
\begin{equation}\label{eq:02}
   U_{ij} = \mu_0 \frac{(\vec{M}_i\vec{M}_j) - 3(\vec{n}\vec{M}_i)(\vec{n}\vec{M}_j)}{r^3},
\end{equation}
where $\mu_0 = 4\pi{\cdot}10^{-7}$~H$/$m is the magnetic vacuum permeability, $r$ is the distance between the endofullerenes, and $\vec{n}$ is the unit vector parallel to the line joining the centers of the endofullerenes.

\subsection{Single magnetic endofullerene at the tip of nanotube}

Let us consider the case where each nanotube of the nanorelay contains a single magnetic endofullerene Fe@C$_\text{20}$ at the tip of the (8,8) nanotube. In this case the magnetic force $F_\text{m}$ of interaction between the nanotubes is given by the result of differentiation of Eq.~\eqref{eq:02} with respect to $2x$. This gives
\begin{equation}\label{eq:03}
   F_\text{m} \equiv \frac{\partial U_{ij}}{\partial(2x)} = N_\text{m}^2f_\text{m} = \frac{6\mu_0M_\text{ef}^2N_\text{m}^2}{(2x)^4} = 24 \frac{\mu_0\mu_\text{B}^2N_\text{m}^2}{x^4},
\end{equation}
where $f_\text{m}$ is the force of interaction between two magnetic moments $M_i = M_j = M_\text{ef} = \text{8}\mu_\text{B}$ which line up in the direction of the magnetic field and $N_\text{m} \leq 1$ is the fraction of the total number $N_\text{ef} = 1$ of the endofullerenes with the magnetic moments parallel to the magnetic field. This number $N_\text{m}$ is determined by the Langevin function $\mathcal{L}(M_\text{ef})$ and depends on the magnetic field induction $B$ and temperature $T$ (see, e.g., Refs.~\citen{White07} and \citen{Borovik05}):
\begin{equation}\label{eq:Langevin}
   N_\text{m} = N_\text{ef}\mathcal{L}(M_\text{ef}) = N_\text{ef}\biggl[\coth\biggl(\frac{M_\text{ef}B}{k_\text{B}T}\biggr) - \frac{k_\text{B}T}{M_\text{ef}B}\biggr],
\end{equation}
where $N_\text{ef} = 1$ is the total number of the endofullerenes inside each nanotube and $k_\text{B}$ is the Boltzmann constant. Since the considered magnetic moments are large we use the Langevin function $\mathcal{L}$ instead of the Brillouin one.

The elastostatic force acting on the carbon nanotube is given by
\begin{equation}\label{eq:04}
   F_\text{s} = \left\{\!\!
\begin{array}{ll}
k(x_\text{o} - x), &\text{for the right nanotube,}\\
k(x - x_\text{o}), &\text{for the left nanotube,}
\end{array}\right.
\end{equation}
where $x$ is the deflection of the nanotube tip, $2x_\text{o}$ is the distance between the axes of the nanotubes in the nanorelay in the absence of a load and $k$ is the bending stiffness coefficient. The macroscopic approach is conventionally used to describe the elastic behavior of nanorelays based on cantilevered nanotubes (see, e.g., Refs.~\citen{Dequesnes02, Kinaret03, Jonsson04}). For a load applied to the nanotube tip the bending stiffness coefficient is given by~\cite{Landau01TVII}
\begin{equation}\label{eq:05}
   k = \frac{3YI_\text{nt}}{L^3},
\end{equation}
where $Y$ is Young's modulus, determined to be approximately 1.2~TPa~\cite{Yakobson01, Eleckij07, Poklonski08CPL187}, $L$ is the nanotube length, $I_\text{nt} = (\pi/4)(R_\text{ex}^4 - R_\text{in}^4)$ is the moment of inertia of the nanotube cross-section, and $R_\text{in}$ and $R_\text{ex}$ are the inner and outer radii, respectively, of the nanotube. The moment of inertia of a single-walled nanotube can be estimated as 
\begin{equation}\label{eq:06}
   I_\text{nt} = \pi R^3\Delta R,
\end{equation}
where $R$ is the nanotube radius and $\Delta R = R_\text{ex} - R_\text{in} \approx{}$0.34~nm is the effective thickness of the single-walled nanotube, which is taken to be equal to the interlayer distance of graphite; $R_\text{ex} = R +{}$0.17~nm and $R_\text{in} = R -{}$0.17~nm. The radius of an ($n,m$) nanotube is determined by the following expression~\cite{Saito98, OConnell06}: $R = (a_\text{C--C}/2\pi)\sqrt{3(n^2 + m^2 + nm)}$, where $a_\text{C--C} ={}$0.142~nm is the C--C bond length in the nanotube. For the (8,8) nanotube $R ={}$0.542 nm.

\begin{figure}
   \begin{center}
   \begin{tabular}{c}
   \includegraphics{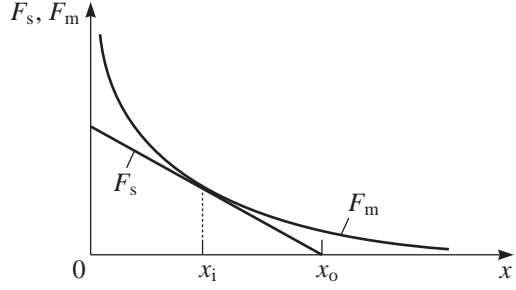}
   \end{tabular}
   \end{center}
   \caption 
   { \label{fig:05} 
The schematic representation of the dependencies of the magnetic force $F_\text{m}$ and the elastostatic force $F_\text{s}$ on the coordinate $x$ of the nanotube tip, $2x_\text{o}$ is the distance between the nanotube tips when magnetic field is absent. The nanorelay is turned on if $F_\text{s} < F_\text{m}$ at any coordinate $x$, i.e., the distance between the nanotube tips should be less than $2x_\text{o}$ for the nanorelay turning on by the magnetic field.} 
   \end{figure}

The nanorelay is turned on by the contact of nanotube tips.
The nanorelay turning on is possible if the magnetic force is larger than the elastostatic force at any deflection $x$ of the nanotube tip. Thus the equalities of magnetic and elastostatic forces $F_\text{s}(x_\text{i}) = F_\text{m}(x_\text{i})$ and their derivatives $\mathrm{d}F_\text{s}/\mathrm{d}x|_{x=x_\text{i}} = \mathrm{d}F_\text{m}/\mathrm{d}x|_{x=x_\text{i}}$ determine the maximum value of the distance $2x_\text{o}$ between the nanotubes (for defined $N_\text{m}$) for which the nanorelay turning on is possible (see Fig.~\ref{fig:05}). Substituting Eqs.~\eqref{eq:03} and \eqref{eq:04} into these equalities, we obtain the following set of equations
\begin{equation}\label{eq:07}
   k = \frac{3}{2} \frac{\mu_0M_\text{ef}^2N_\text{m}^2}{x_\text{i}^5},\quad
   k(x_\text{o} - x_\text{i}) = \frac{3}{8} \frac{\mu_0M_\text{ef}^2N_\text{m}^2}{x_\text{i}^4}.
\end{equation}
Hence it follows that
\begin{equation}\label{eq:07'}
   x_\text{i} = \frac{4}{5} x_\text{o},\quad
   N_\text{m} = \frac{32}{375} \frac{x_\text{o}^{5/2}}{M_\text{ef}} \frac{\sqrt{30k}}{\sqrt{\mu_0}},
\end{equation}
where $x_\text{i}$ is the coordinate of the nanotube tip when $F_\text{m} = F_\text{s}$  and $x_\text{o} \ll L$. 

The amplitude of thermal oscillations of the nanotubes has to be small enough to preclude the possibility of spontaneous closure of the nanorelay in the absence of the magnetic field. The nanorelay cannot operate in the case of the average thermal oscillation amplitude larger than $x_\text{o}^\text{th} - x_\text{o}^\text{min}$, where $2x_\text{o}^\text{min} = 2R + \Delta R$ is the distance between the nanotubes which corresponds to their contact. For this condition one can use an estimate $k(x_\text{o}^\text{th} - x_\text{o}^\text{min})^2/2 > k_\text{B}T/2$, where $2x_\text{o}^\text{th}$ is the minimal distance between the nanotubes limited by thermal oscillations. More accurately, by taking the expression for the average thermal oscillation amplitude~\cite{Treacy96}, this condition is given by
\begin{equation}\label{eq:08}
   (x_\text{o}^\text{th} - x_\text{o}^\text{min})^2 > 0.4243 \frac{\pi L^3k_\text{B}T}{4YI_\text{nt}},
\end{equation}
where $k_\text{B}T$ is thermal energy.

   \begin{figure}[!t]
   \begin{center}
   \begin{tabular}{c}
   \includegraphics{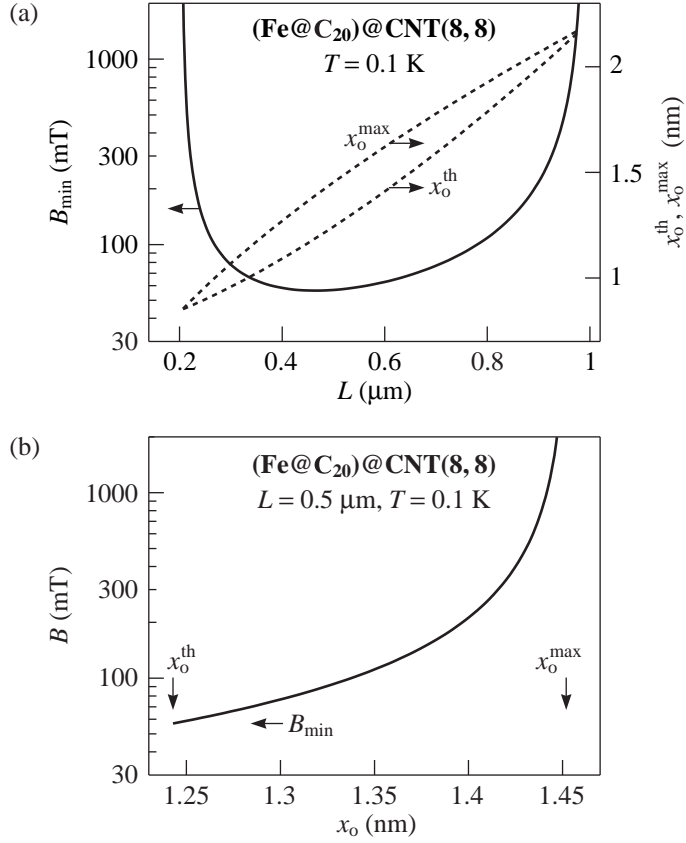}
   \end{tabular}
   \end{center}
   \caption 
   { \label{fig:06} 
(a) The dependence of the minimal magnetic field induction $B_\text{min}$ on the nanotube length $L$ for the nanorelay operated at $T ={}$0.1~K and based on the (8,8) carbon nanotubes  (left axis). Each nanotube contains single magnetic endofullerene Fe@C$_\text{20}$ with $M_\text{ef} = \text{8}\mu_\text{B}$ at the tip. The dependences of half of the minimal ($2x_\text{o}^\text{th}$) and maximal ($2x_\text{o}^\text{max}$) distances between the nanotubes for which the nanorelay operation is possible on the nanotube length $L$ (right axis); $x_\text{o}^\text{min} ={}$0.712~nm. (b)~The dependence of the induction $B$ necessary for the nanorelay operation with respect to the distance $2x_\text{o}$ between the nanotubes for the nanotube length $L ={}$0.5~$\mu$m. Arrows indicate for this $L$ the values of $x_\text{o}^\text{th}$, $x_\text{o}^\text{max}$ and $B_\text{min}$ shown in Fig.~\ref{fig:06}(a).}
   \end{figure} 

We obtain from Eq.~\eqref{eq:07'} at $N_\text{m} = 1$ the maximal distance $2x_\text{o}^\text{max}$ between the nanotubes for the nanorelay turning on for a given nanotube lengths:
\begin{equation}\label{eq:x_omax}
   x_\text{o}^\text{max} = \biggl(\frac{375M_\text{ef}}{32} \frac{\sqrt{\mu_0}}{\sqrt{30k}}\biggr)^{2/5}.
\end{equation}
Substituting Eq.~\eqref{eq:07'} and $x_\text{o} = x_\text{o}^\text{th}$ (according to Eq.~\eqref{eq:08}) into Eq.~\eqref{eq:Langevin}, the minimal magnetic field induction $B_\text{min}$ which is necessary for the nanorelay turning on can be numerically calculated. The dependence of the minimal magnetic field induction $B_\text{min}$ on the nanotube length $L$ for the nanorelay based on the (8,8) nanotubes operated at $T ={}$0.1~K is shown in Fig.~\ref{fig:06}(a). Figure~\ref{fig:06}(a) also shows the dependence of half of the minimal ($2x_\text{o}^\text{th}$) and maximal ($2x_\text{o}^\text{max}$) distances (according to Eqs.~\eqref{eq:08} and  \eqref{eq:x_omax}, respectively) between the nanotubes for which the nanorelay operation is possible with respect to the nanotube length $L$ corresponding to the case under consideration. Figure~\ref{fig:06}(b) shows the dependence of the induction $B$ necessary for the nanorelay operation on the distance $2x_\text{o}$ between the nanotubes for nanotube length $L = 0.5$~$\mu$m.

The measurements of van der Waals forces between the caps of the nanotubes about 1~nm in diameter gave the maximal value $\approx{}$1--2~nN~\cite{Nagataki09}. For the considered case of the interaction between the lateral surfaces of nearly parallel nanotubes the van der Waals forces should be significantly larger. Whereas for the (8,8) nanotube 0.5~$\mu$m in length the bending stiffness coefficient is equal to $k ={}$5~nN{/}mm, and the elastostatic force for the deflection of the nanotube tip, $x_\text{o} ={}$1.2~nm, is equal to $F_\text{s} \approx kx_\text{o} = 6{\cdot}10^{-6}$~nN. Thus the van der Waals forces are significantly larger than the elastostatic forces. The nanorelay turning off can be achieved by applying an electrostatic potential (without current) to the same wire used for the nanorelay turning on~\cite{Jang05}.

The switching time $\tau$ of the nanorelay can not be considerably less than the period of free bending vibrations of the nanotube. The frequencies of the free bending vibrations of the nanotube are given by~\cite{Treacy96}:
\begin{equation}
   \omega_s = 2\pi f_s = \frac{\beta_s^2}{2L^2} \sqrt{\frac{Y(R_\text{ex}^2 + R_\text{in}^2)}{\rho}},
\end{equation}
where $\beta_s$ for a given mode $s$ is found from the solution to the equation $\cos\beta_s\cosh\beta_s + 1 = 0$; $\beta_0 \approx{}$1.8751 for the fundamental mode, $\rho \approx{}$2.15~g$/$cm$^3$ is the density of the material in the nanotube wall. For the (8,8) nanotube ($R_\text{ex} ={}$0.712~nm, $R_\text{in} ={}$0.372~nm) of the length $L ={}$0.5~$\mu$m we estimate:
\begin{equation}
   \tau \approx \frac{1}{f_0} = 2\pi \frac{2L^2}{\beta_0^2} \sqrt{\frac{\rho}{Y(R_\text{ex}^2 + R_\text{in}^2)}} \approx \text{50 ns}.
\end{equation}
The induction $B$ of the magnetic field induced by the current $I$ at the distance $d$ from the wire is~\cite{Benenson02}:
\begin{equation}\label{eq:10}
   B = \frac{\mu_0I}{2\pi d}.
\end{equation}
To turn on the nanorelay by applying the magnetic field with the induction of 60~mT (for the (8,8) nanotube of 0.5~$\mu$m in length, see Fig.~\ref{fig:06}) it is necessary to use the current $I ={}$300~mA passing through the wire \emph{3} (see Fig.~\ref{fig:04}), which is positioned at distance $d ={}$1~$\mu$m from nanotubes, for the time interval somewhat longer than $\tau \approx{}$50~ns.

\subsection{Nanotube fully filled with magnetic endofullerenes}

Let us consider the magnetic interaction between two parallel peapod nanotubes of the length $L$ fully filled with magnetic endofullerenes Fe@C$_\text{20}$. The number $N_\text{m} \gg 1$ of the endofullerenes with magnetic moments parallel to the magnetic field is determined by the total number of encapsulated magnetic endofullerenes $N_\text{ef} \gg 1$, the magnetic field induction and temperature. By summing up the forces $\vec{F}_{ij} \equiv \nabla U_{ij}$, where $U_{ij}$ is given by Eq.~\eqref{eq:02}, over all pairs of such magnetic dipoles, it can be easily shown that for the case where the distance $2x$ between the nanotubes is considerably less than the nanotube length $L$, $2x \ll L$ (i.e., neglecting the tube edge effects), the magnetic force of interaction between the nanotubes takes the form 
\begin{equation}\label{eq:11}
   F_\text{m} = \sum_{i = 1}^{N_\text{m}} \sum_{j = 1}^{N_\text{m}}\frac{\partial U_{ij}}{\partial(2x)} \approx 4 \frac{\mu_0M_\text{ef}^2N_\text{m}^2}{L(2x)^3} = 32 \frac{\mu_0\mu_\text{B}^2N_\text{m}^2}{Lx^3},
\end{equation}
where $N_\text{m}$ is the number of the endofullerenes with magnetic moments parallel to the magnetic field inside each nanotube.

We assume that the magnetic endofullerenes inside the nanotubes form a superparamagnetic phase. 
In this case the number $N_\text{m}$ of the endofullerenes with magnetic moments parallel to the magnetic field is determined by the Langevin function (see, e.g., Refs.~\citen{White07} and \citen{Borovik05}):
\begin{equation}\label{eq:09}
   N_\text{m} = N_\text{ef}\mathcal{L}(N_\text{ef}M_\text{ef}) = N_\text{ef}\biggl[\coth\biggl(\frac{N_\text{ef}M_\text{ef}B}{k_\text{B}T}\biggr) - \frac{k_\text{B}T}{N_\text{ef}M_\text{ef}B}\biggr],
\end{equation}
where $B$ is the magnetic field induction, $N_\text{ef}$ is the number of the magnetic endofullerenes inside the nanotube, and $N_\text{ef}M_\text{ef}$ is the magnetic moment corresponding to the magnetic saturation. 

For homogeneous distribution of a load along the nanotube length the bending stiffness coefficient is given by~\cite{Landau01TVII}
\begin{equation}\label{eq:12}
   k = \frac{8YI_\text{r}}{L^3},
\end{equation}
where the moment of inertia of the cross-section of the nanotube fully filled with the endofullerenes is
\begin{equation}\label{eq:13}
   I_\text{r} = \pi R^4/4.
\end{equation}
Here $R = 3na_\text{C--C}/2\pi$ is the radius of the ($n,n$) nanotube and $a_\text{C--C} ={}$0.142~nm.

The nanorelay turning on is possible if the magnetic force $F_\text{m}$ is larger than the elastostatic force $F_\text{s}$ at any deflection $x$ of the nanotube tip. Thus the equalities of magnetic and elastostatic forces $F_\text{s}(x_\text{i}) = F_\text{m}(x_\text{i})$ and their derivatives $\mathrm{d}F_\text{s}/\mathrm{d}x|_{x=x_\text{i}} = \mathrm{d}F_\text{m}/\mathrm{d}x|_{x=x_\text{i}}$ determine the maximum value of the distance $2x_\text{o}$ between the nanotubes (for defined $N_\text{m}$) for which the nanorelay turning on is possible (see Fig.~\ref{fig:05}). Substituting expressions \eqref{eq:04} and \eqref{eq:11} into these equalities, we obtain the following set of equations
\begin{equation}\label{eq:14}
   k = \frac{3}{2} \frac{\mu_0M_\text{ef}^2N_\text{m}^2}{Lx_\text{i}^4},\quad
   k(x_\text{o} - x_\text{i}) = \frac{1}{2} \frac{\mu_0M_\text{ef}^2N_\text{m}^2}{Lx_\text{i}^3}.
\end{equation}
Hence it follows that
\begin{equation}\label{eq:14'}
   x_\text{i} = \frac{3}{4}x_\text{o},\quad
   N_\text{m} = \frac{3}{16} \frac{x_\text{o}^2}{M_\text{ef}} \frac{\sqrt{6kL}}{\sqrt{\mu_0}}.
\end{equation}

   \begin{figure}[!t]
   \begin{center}
   \begin{tabular}{c}
   \includegraphics{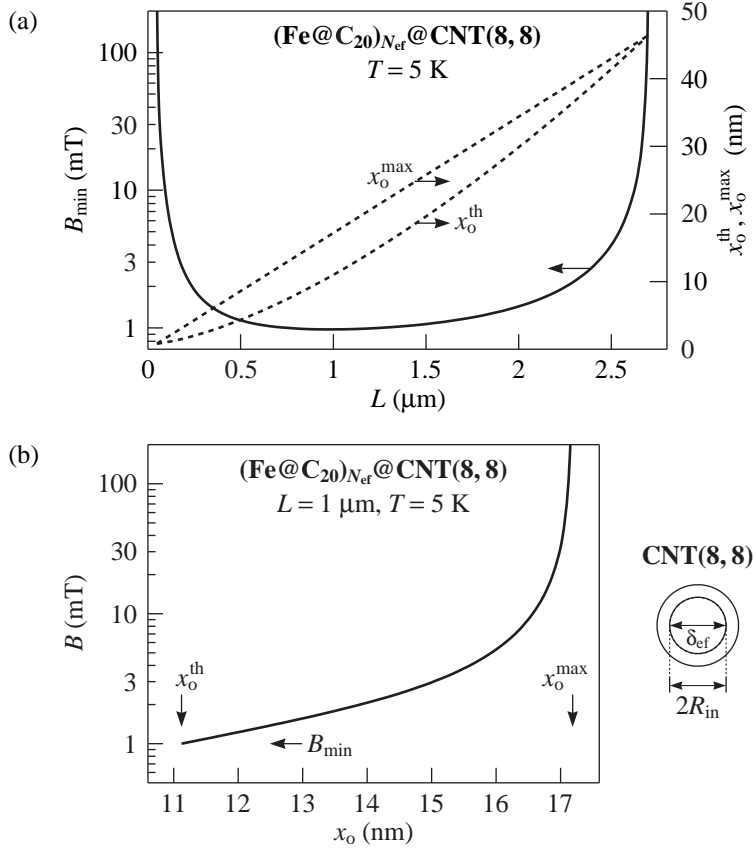}
   \end{tabular}
   \end{center}
   \caption 
   { \label{fig:07} 
(a) The dependence of the minimal magnetic field induction $B_\text{min}$ on the nanotube length $L$ for the nanorelay operated at $T ={}$5~K and based on the (8,8) carbon nanotubes fully filled with the magnetic endofullerenes Fe@C$_\text{20}$ (left axis). Here $N_\text{ef} \gg 1$ is the number of the endofullerenes in the fully filled (8,8) nanotube. The dependences of half of the minimal ($2x_\text{o}^\text{th}$) and maximal ($2x_\text{o}^\text{max}$) distances between the nanotubes for which the nanorelay operation is possible on the nanotube length $L$ (right axis); $x_\text{o}^\text{min} ={}$0.712~nm. (b)~The dependence of the induction $B$ necessary for the nanorelay operation with respect to the distance $2x_\text{o}$ between the nanotubes for the nanotube length $L ={}$1~$\mu$m. Arrows indicate the values of $x_\text{o}^\text{th}$, $x_\text{o}^\text{max}$ and $B_\text{min}$ shown in Fig.~\ref{fig:07}(a). The cross-section of the (8,8) nanotube ($2R_\text{in} \approx \text{0.75}$~nm) filled with Fe@C$_\text{20}$ is shown in the right part of the figure; here $\delta_\text{ef} \approx (\text{0.42} + \text{0.34})$~nm and $M_\text{ef} = \text{8}\mu_\text{B}$.}
   \end{figure} 

The expression for the average thermal oscillation amplitude in the considered case of homogeneous distribution of a load along the nanotube length takes the form (see Eq.~\eqref{eq:08})
\begin{equation}\label{eq:15}
   (x_\text{o}^\text{th} - x_\text{o}^\text{min})^2 = \text{0.4243}\, \frac{3}{8} \frac{\pi L^3k_\text{B}T}{4YI_\text{r}},
\end{equation}
where the coefficient $3/8$ is introduced into this equation because of the different bending stiffness coefficients in the cases of distributed and lumped load (cf{.} Eqs.~\eqref{eq:05} and \eqref{eq:12}); $x_\text{o}^\text{min} = R +{}$0.17~nm.

   \begin{figure}[!t]
   \begin{center}
   \begin{tabular}{c}
   \includegraphics{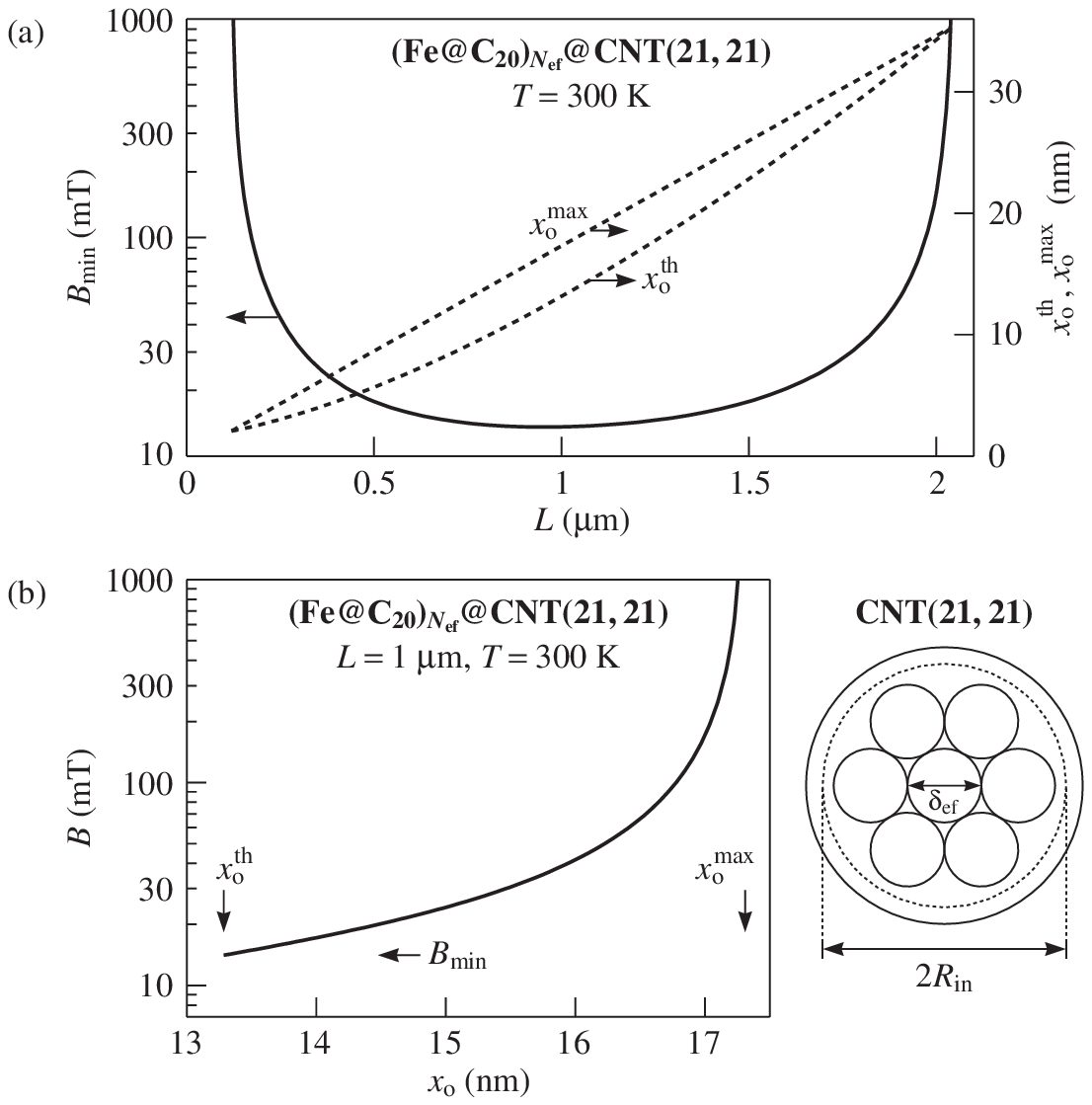}
   \end{tabular}
   \end{center}
   \caption 
   { \label{fig:08} 
(a) The dependence of the minimal magnetic field induction $B_\text{min}$ on the nanotube length $L$ for the nanorelay operated at $T ={}$300~K and based on the (21,21) carbon nanotubes fully filled with the magnetic endofullerenes Fe@C$_\text{20}$ (left axis). Here $N_\text{ef} \gg 1$ is the number of the endofullerenes in the fully filled (21,21) nanotube. The dependences of half of the minimal ($2x_\text{o}^\text{th}$) and maximal ($2x_\text{o}^\text{max}$) distances between the nanotubes for which the nanorelay operation is possible on the nanotube length $L$ (right axis); $x_\text{o}^\text{min} ={}$1.594~nm. (b)~The dependence of the induction $B$ necessary for the nanorelay operation with respect to the distance $2x_\text{o}$ between the nanotubes for the nanotube length $L ={}$1~$\mu$m. Arrows indicate the values of $x_\text{o}^\text{th}$, $x_\text{o}^\text{max}$ and $B_\text{min}$ shown in Fig.~\ref{fig:08}(a). The cross-section of the (21,21) nanotube ($2R_\text{in} \approx \text{2.85}$~nm) filled with Fe@C$_\text{20}$ is shown in the right part of the figure; here $\delta_\text{ef} \approx (\text{0.42} + \text{0.34})$~nm.}
   \end{figure} 

The number $N_\text{ef}$ of the endofullerenes in the fully filled (8,8) nanotube we estimate as
\begin{equation}\label{eq:16}
   N_\text{ef} = \frac{L}{\delta_\text{ef}},
\end{equation}
where $\delta_\text{ef} \approx (\text{0.42} + \text{0.34})$\,nm is the distance between the centers of the adjacent spherical endofullerenes. Here 0.42~nm is the diameter of endofullerene Fe@C$_\text{20}$, and 0.34~nm is the distance between the surfaces of adjacent endofullerenes which we assume to be equal to the interlayer distance of graphite.

   \begin{figure}[!t]
   \begin{center}
   \begin{tabular}{c}
   \includegraphics{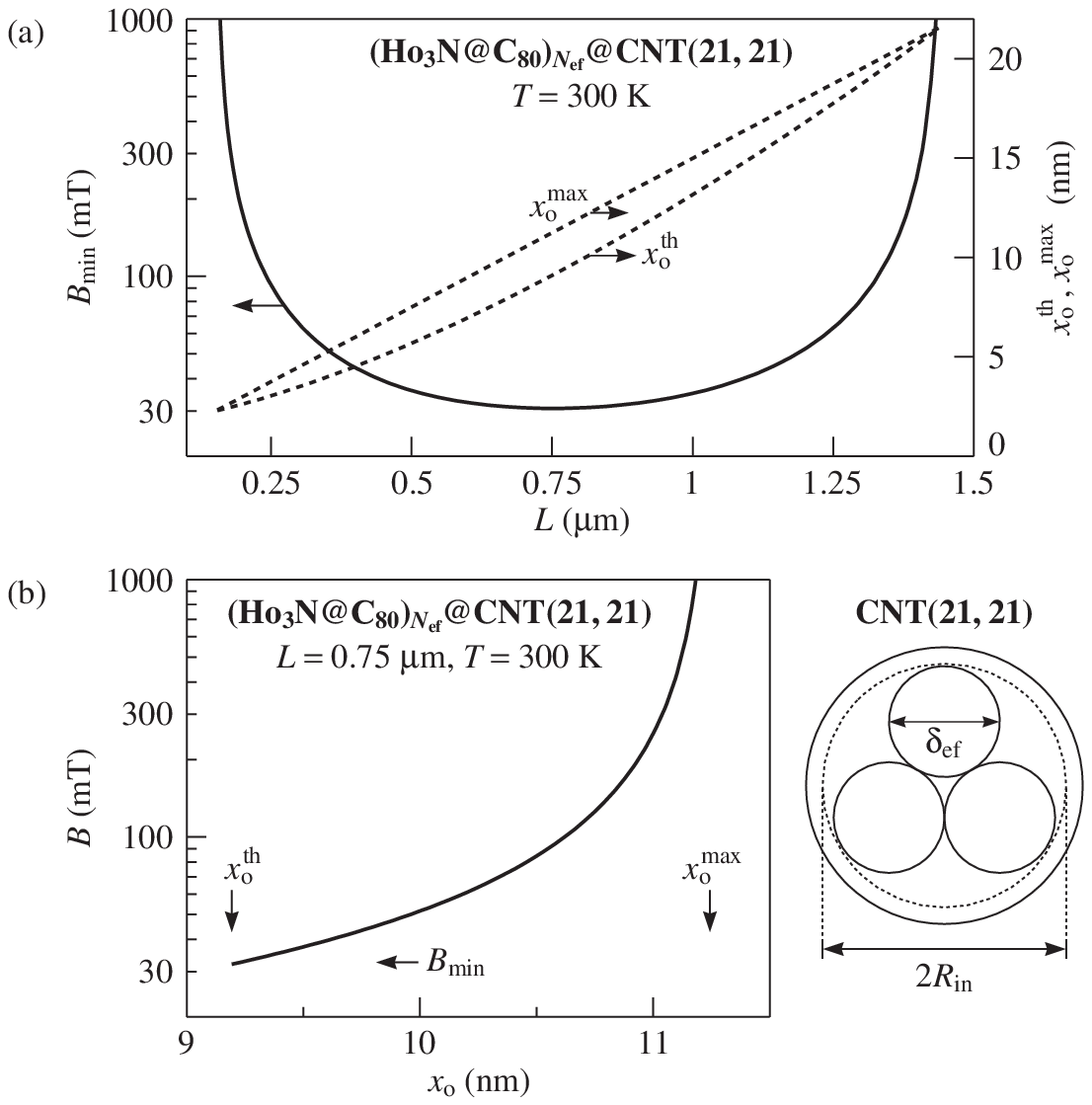}
   \end{tabular}
   \end{center}
   \caption 
   { \label{fig:09} 
(a) The dependence of the minimal magnetic field induction $B_\text{min}$ on the nanotube length $L$ for the nanorelay operated at $T ={}$300~K and based on the (21,21) carbon nanotubes fully filled with the magnetic endofullerenes (Ho$_\text{3}$N)@C$_\text{80}$ (left axis). Here $N_\text{ef} \gg 1$ is the number of the endofullerenes in the fully filled (21,21) nanotube. The dependences of half of the minimal ($2x_\text{o}^\text{th}$) and maximal ($2x_\text{o}^\text{max}$) distances between the nanotubes for which the nanorelay operation is possible on the nanotube length $L$ (right axis); $x_\text{o}^\text{min} ={}$1.594~nm. (b)~The dependence of the induction $B$ necessary for the nanorelay operation with respect to the distance $2x_\text{o}$ between the nanotubes for the nanotube length $L ={}$0.75~$\mu$m. Arrows indicate the values of $x_\text{o}^\text{th}$, $x_\text{o}^\text{max}$ and $B_\text{min}$ shown in Fig.~\ref{fig:09}(a). The cross-section of the (21,21) nanotube ($2R_\text{in} \approx \text{2.85}$~nm) filled with (Ho$_\text{3}$N)@C$_\text{80}$ is shown in the right part of the figure; here $\delta_\text{ef} \approx (\text{0.8} + \text{0.34})$~nm and $M_\text{ef} = \text{21}\mu_\text{B}$.}
   \end{figure}

From Eq.~\eqref{eq:14'} at $N_\text{m} = N_\text{ef}$ the maximal distance $2x_\text{o}^\text{max}$ between the nanotubes for the nanorelay turning on for a given nanotube lengths was obtained
\begin{equation}\label{eq:xomaxfilled}
   x_\text{o}^\text{max} = \biggl(\frac{16N_\text{ef}M_\text{ef}}{3} \frac{\sqrt{\mu_0}}{\sqrt{6kL}}\biggr)^{\!1/2}.
\end{equation}

Substituting Eq.~\eqref{eq:14'} and $x_\text{o} = x_\text{o}^\text{th}$ (according to Eq.~\eqref{eq:15}) into Eq.~\eqref{eq:09}, the minimal magnetic field induction $B_\text{min}$ which is necessary for the nanorelay turning on can be numerically calculated. The dependence of the minimal magnetic field induction $B_\text{min}$ on the nanotube length $L$ for the nanorelay based on the (8,8) nanotubes operated at $T ={}$5~K is shown in Fig.~\ref{fig:07}(a). For the (8,8) nanotubes there is no solution of Eqs.~\eqref{eq:09}, \eqref{eq:14'} and \eqref{eq:15} at room temperature (it follows from Eqs.~\eqref{eq:14'} and \eqref{eq:15} that the number $N_\text{m}$ of moments aligned with field necessary for the nanorelay turning on is larger than the number $N_\text{ef}$ of endofullerenes which the nanotube can contain (see Eq.~\eqref{eq:16}). 

When we use the (21,21) nanotubes the nanorelay can operate at room temperature (see Fig.~\ref{fig:08}). We assume that the endofullerenes are arranged in the hexagonal dense packing in the nanotube cross-section and in the form of seven chains along the nanotube axis (see the inset in Fig.~\ref{fig:08}(b)). 
It should be noted that for the nanotube of the large diameter the number $N_\text{ef}$ of the endofullerenes in the fully filled nanotube one can estimate as
\begin{equation}\label{eq:17}
   N_\text{ef} = \mathop{\mathrm{Int}}\biggl[\frac{\pi\sqrt{3}}{6} \frac{\pi R_\text{in}^2}{S_\text{ef}}\biggr] \frac{L}{\delta_\text{ef}},
\end{equation}
where $\mathop{\mathrm{Int}}[N]$ is the integer part of the number $N$, the area $S_\text{ef}$ occupied by one endofullerene we estimate as $S_\text{ef} \approx \pi\delta_\text{ef}^2/4$ for the distance $\delta_\text{ef} \approx (\text{0.42} + \text{0.34})$\,nm between centers of the neighbor endofullerenes Fe@C$_\text{20}$; coefficient $\pi\sqrt{3}/6$ is taken for the hexagonal dense packing of circles in the plain~\cite{Sloane84}. If the endofullerenes inside a nanotube are polymerized and form covalent bonds, then $\delta_\text{ef}$ will be smaller. Note that $R_\text{in} = R -{}$0.17~nm and $R = 3na_\text{C--C}/2\pi$ is the radius of the ($n,n$) nanotube. Formula \eqref{eq:17} gives the asymptotic upper bound of $N_\text{ef}$ for $\pi R_\text{in}^2/S_\text{ef} \gg 1$.

In Ref.~\citen{Wolf05} the existence of the magnetic endofullerene (Ho$_\text{3}$N)@C$_\text{80}$ with the total magnetic moment of the molecule $M_\text{ef} = \text{21}\mu_\text{B}$ was shown experimentally. Our calculations of operational characteristics of the nanorelay made of (21,21) nanotubes filled with (Ho$_\text{3}$N)@C$_\text{80}$ at room temperature are presented in Fig.~\ref{fig:09}. We assume that the endofullerenes form three chains along the nanotube axis (see inset in Fig.~\ref{fig:09}(b)). The distance between centers of the neighbor endofullerenes (Ho$_\text{3}$N)@C$_\text{80}$ is $\delta_\text{ef} \approx (\text{0.8} + \text{0.34})$~nm. Here 0.8~nm is the diameter of endofullerene (Ho$_\text{3}$N)@C$_\text{80}$, and 0.34~nm is the distance between the surfaces of adjacent endofullerenes which we assume to be equal to the interlayer distance of graphite.


\section{Conclusion}

In this paper, we applied the DFT to calculate the structural and energy characteristics of the smallest magnetic endofullerene Fe@C$_\text{20}$. The ground state of Fe@C$_\text{20}$ was found to be the septet state, and the magnetic moment of Fe@C$_\text{20}$ was estimated to be $\text{8}\mu_\text{B}$. The characteristics of the equilibrium position and motion of this endofullerene inside the (8,8) nanotube were studied in the framework of the semiempirical approach. The calculated barriers for the motion and rotation show that the endofullerene freely moves and rotates inside the nanotube at room temperature.

The scheme of the magnetic nanorelay based on cantilevered nanotubes filled with magnetic endofullerenes is elaborated. The proposed nanorelay is turned on as a result of bending of the nanotubes by a magnetic force. Operational characteristics of the nanorelay based on the (8,8) and (21,21) nanotubes fully filled with the proposed here endofullerenes Fe@C$_\text{20}$ and on (21,21) nanotube fully filled with observed (Ho$_\text{3}$N)@C$_\text{80}$ endofullerenes are calculated. It is shown that the nanorelay based on the (21,21) nanotubes fully filled with the endofullerenes Fe@C$_\text{20}$ or (Ho$_\text{3}$N)@C$_\text{80}$ can operate at room temperature for the nanotubes of about 1~$\mu$m in length.

The proposed nanorelay can be used also for the measurements of magnetic moments of endofullerenes or other magnetic molecules inside single-walled carbon nanotubes and for investigation of elastic properties of nanotubes. 

\acknowledgments 
This work has been partially supported by the BFBR (Grants F08R-061 and F08VN-003) and RFBR (Grants 08-02-00685 and 08-02-90049-Bel).


\bibliography{NanoReedSwitch2009}   
\bibliographystyle{spiejour}   


\newpage
\vspace{2ex}\noindent\textbf{Nikolai A. Poklonski} is a professor at the Belarusian State University. He received his Dr. Sci. degree in semiconductor physics from Belarusian State University in 2001. He is the author of more than 170 journal papers. His current research interests include semiconductor physics and electronics, radiospectroscopy of diamond, physics of carbon nanosystems.

\vspace{2ex}\noindent\textbf{Eugene F. Kislyakov} is a senior scientific researcher at the Belarusian State University. He received his degrees in physics from Moscow State University (PhD in nuclear physics 1982). He is the author of more than 50 journal papers. His current research interests include carbon nanosystems.

\vspace{2ex}\noindent\textbf{Sergey A. Vyrko} is a senior scientific researcher at the Belarusian State University. He received his PhD in semiconductor physics from Belarusian State University in 2002. He is the author of more than 25 journal papers. His current research interests include physics of heavily doped semiconductors, hopping conductivity, phase transitions in semiconductors, physics of fullerenes and carbon nanostructures.

\vspace{2ex}\noindent\textbf{Nguyen Ngoc Hieu} is a scientific researcher at the Institute of Physics and Electronics. He received his PhD in semiconductor physics from Belarusian State University in 2009. He is the author of more than 10 journal papers. His current research interests include physics of carbon nanostructures.

\vspace{2ex}\noindent\textbf{Oleg N. Bubel'} is an associate professor at the Belarusian State University. He received his PhD in organic chemistry from Belarusian State University in 1972. He is the author of more than 170 journal papers. He is an expert in quantum chemistry. His current research interests include carbon nanosystems.

\vspace{2ex}\noindent\textbf{Andrei I. Siahlo} is a senior scientific researcher at the Belarusian State University. He received his PhD in semiconductor physics from Belarusian State University in 2000. He is the author of more than 50 publications. His current research interests include physics of semiconductor devices, and physics of carbon nanostructures.

\vspace{2ex}\noindent\textbf{Irina V. Lebedeva} is a PhD student at Moscow Institute of Physics and Technology. She is the co-author of 5 journal papers. Her current research interests include atomistic modeling and physics of carbon nanostructures.

\vspace{2ex}\noindent\textbf{Andrey A. Knizhnik} is a senior scientific researcher at the Russian Research State ``Kurchatov Institute'' and Kintech Lab Ltd. He received his degrees in physics from Moscow Institute of Physics and Technology (PhD in plasma physics 2002). He is the author of more than 35 journal papers. His current research interests include atomistic and first-principles modeling, multiscale modeling, physics of nanomaterials and nanosystems.

\vspace{2ex}\noindent\textbf{Andrey M. Popov} is a senior scientific researcher at the Institute of Spectroscopy. He received his degrees in physics from Institute of Spectroscopy (PhD in theoretical physics 2007). He is the author of more than 50 journal papers. His current research interests include nanodevices based on carbon nanotubes and graphene, phase transitions in nanosystems.

\vspace{2ex}\noindent\textbf{Yurii E. Lozovik} is a head of laboratory of Institute of Spectroscopy RAS, professor of Moscow Institute of Physics and Technology. He is the author of more than 500 publications including reviews and collective monographs. His current research interests include nanophysics and nanooptics.

\end{document}